\documentclass[prb,showpacs,superscriptaddress,twocolumn,citeautoscript]{revtex4}
\usepackage{graphicx}
\usepackage{float}
\usepackage{amsmath}
\usepackage{amssymb}
\usepackage{color}
\usepackage{subfigure}

\newcommand{\rmd}{{\rm d}}

\begin{document}

\title{Self-consistent calculation of electric potentials in Hall devices.}
\author{Tobias~Kramer}
\email{tobias.kramer@mytum.de}
\affiliation{Institute for Theoretical Physics, University of Regensburg, 93040 Regensburg, Germany}
\affiliation{Department of Physics, Harvard University, Cambridge, MA~02138, USA}
\author{Viktor Krueckl}
\affiliation{Institute for Theoretical Physics, University of Regensburg, 93040 Regensburg, Germany}
\author{Eric~J.~Heller}
\affiliation{Department of Physics, Harvard University, Cambridge, MA~02138, USA}
\affiliation{Department of Chemistry and Chemical Biology, Harvard University, Cambridge, MA~02138, USA}
\author{Robert~E.~Parrott}
\affiliation{Department of Physics, Harvard University, Cambridge, MA~02138, USA}
\affiliation{School of Engineering and Applied Science, Harvard University, Cambridge, MA~02138, USA}

\begin{abstract}
Using a first-principles classical many-body simulation of a Hall bar, we study the necessary conditions for the formation of the Hall potential: (i) Ohmic contacts with metallic reservoirs, (ii) electron-electron interactions, and (iii) confinement to a finite system. By propagating thousands of interacting electrons over million time-steps we capture the build-up of the self-consistent potential. The microscopic model sheds light on the  the current injection process and directly links the Hall effect to specific boundary conditions at the particle reservoirs.
\end{abstract}


\maketitle

\section{Introduction}

The calculation of self-consistent potentials of an interacting many-body system is of paramount importance for understanding the transport phenomena occurring in nanostructures and to develop realistic semiconductor device-models reaching from the classical to the quantum regime \cite{Vasileska2008a}. Ground state density functional theory is not always suitable to treat transport phenomena in semiconductors \cite{Koentopp2008a}, and time-dependent density functional theory is still limited to rather small systems (less than hundreds of electrons). On the other hand, classical theories for interacting electronic systems are well established, i.e.\  Thomas-Fermi screening and its extensions \cite{Pittalis2009a} have been successfully used to obtain the effective mean-field potential of electrons in a semiconductor device.
\begin{figure}[b]
\centering
\subfigure[]{
\includegraphics[width=0.47\textwidth]{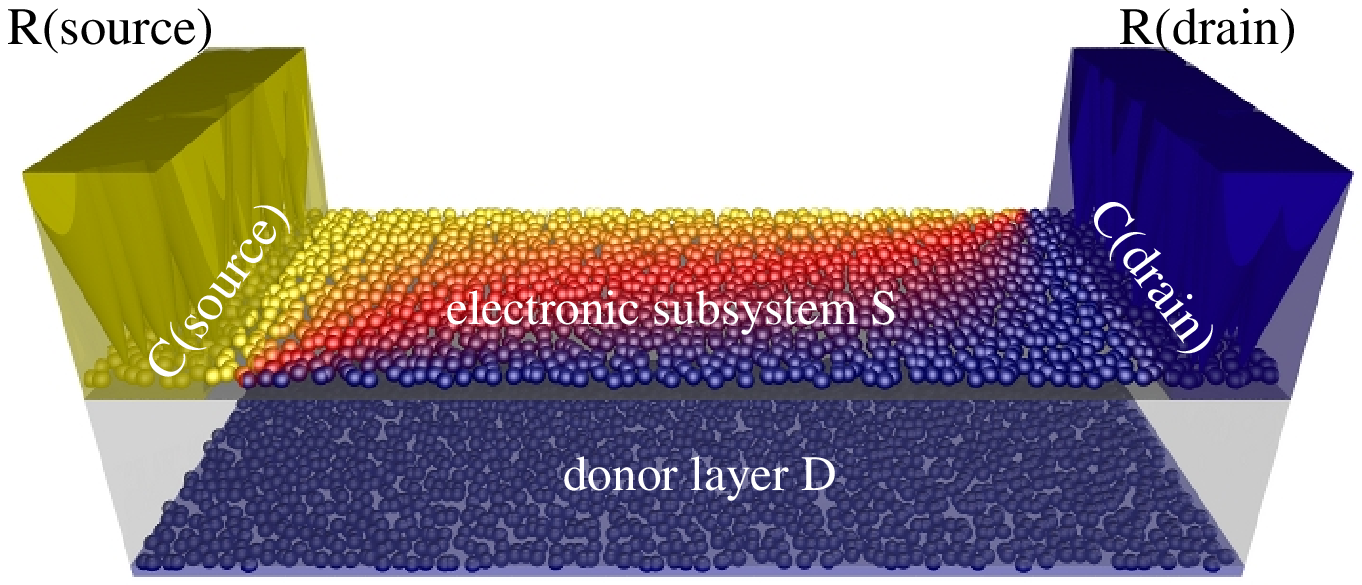}\label{fig:System}
}
\subfigure[]{
\includegraphics[width=0.47\textwidth]{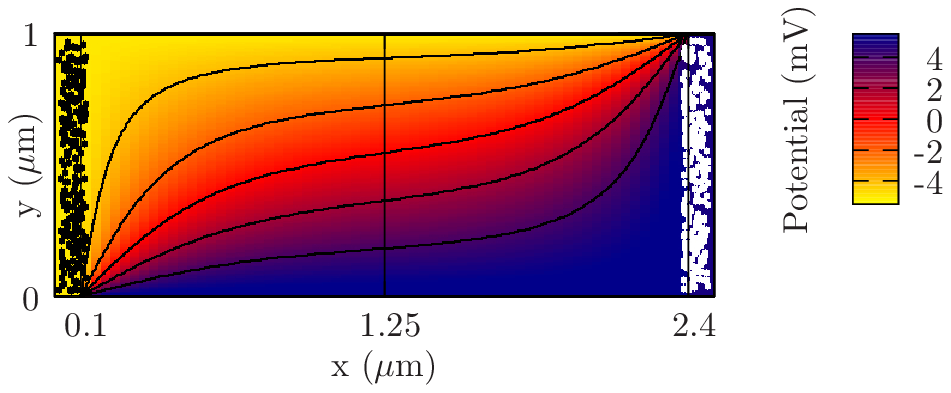}\label{fig:LaplaceHallPotential}
}
\subfigure[]{
\includegraphics[width=0.47\textwidth]{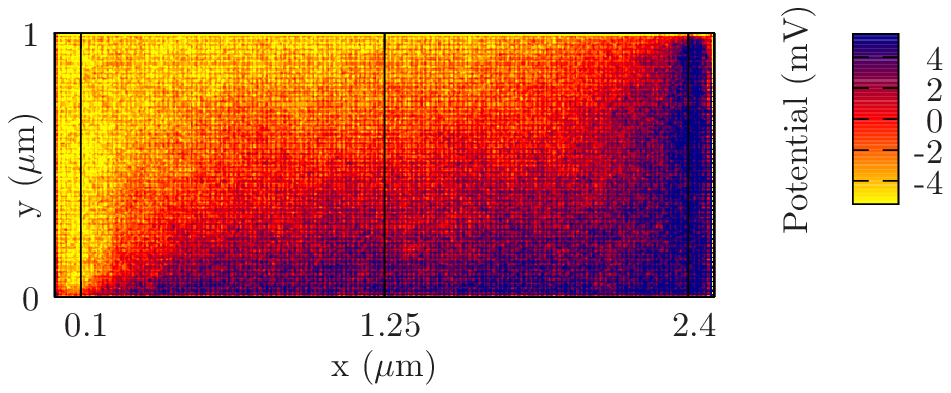}\label{fig:AveragePotential}
}
\caption{(color online)
(a) System setup. The metallic reservoirs R supply and remove electrons through the contact stripes C to/from the two-dimensional electronic subsystem S, which sits above the positively charged donor layer D.
(b) Hall potential obtained from the conformal mapping method, where electrons enter the Hall device from the left source contact and drift to the right one. The contacts are metallic equipotential surfaces. The black lines denote equipotential lines at $-4$,$-2$,$0$,$2$,$4$~mV. The dots at the left and right sides denote the positions of 500 contact points C$_i$ used in the numerical simulation to model the Ohmic contact connecting the metallic reservoir with the electronic subsystem-layer.
(c) Time-averaged potential obtained from the microscopic particle-simulation described in Sect.~\ref{sec:result}.
}
\end{figure}
The presence of a stationary current through a device requires  implementing an open-system, where the contacts provide sources and sinks for electrons. The contacts form an indispensable part of the simulated system. The methods described above are only partially able to address this situation, even for a purely classical system \cite{Albareda2009a}.  An interesting example is the classical Hall effect which gives rise to an unusual potential-theory problem \cite{Moelter1998a}. In a Hall bar, the voltages at the source and drain contacts are externally given and the current flow has to be calculated under the condition that the electrons do not cross the sample edges. These constraints on the current in the device and the voltages at the contacts have to be satisfied simultaneously. The resulting Hall potential (Fig.~\ref{fig:LaplaceHallPotential}) has been obtained via conformal mapping \cite{Wick1954a,Rendell1981a}. The conformal map yields a unidirectional current flow, without counterpropagating currents along the edges. Instead a diagonal current flow from one corner of the device to the opposite corner emerges. Recently within the non-equillibrium network approach similar potentials were obtained \cite{Oswald2005a,Uiberacker2009a}.

Mainly due to the lack of computational feasibility, a first-principles, microscopic model of the Hall effect has been lacking. But only such a model allows us to study and change important parameters (density, donor-layer distance, device geometry, gates, etc.) and to finally understand which characteristics of a device, microscopic forces, and boundary conditions are necessary to generate the experimentally observed Hall potentials \cite{Knott1995a}. To achieve this step, we have developed a new computational approach utilizing graphics processing units and adapted many-body algorithms used to study galaxy formation to nanodevices. In addition, our particle-based simulations highlight the dynamical nature of electron transport in a device, where we study the time-dependent build up of the self-consistent particle distribution.

The manuscript is organized as follows: in Sect.~\ref{sec:model} we describe the general idea of the particle-based simulation scheme and give the details of our method in Sect~\ref{sec:detail}. Results are discussed in Sects.~\ref{sec:result}, \ref{sec:boundary} and finally we present conclusion and an outlook for future research in Sect.~\ref{sec:conclusion}.

\begin{figure}[t]
\begin{center}
\includegraphics[width=0.47\textwidth]{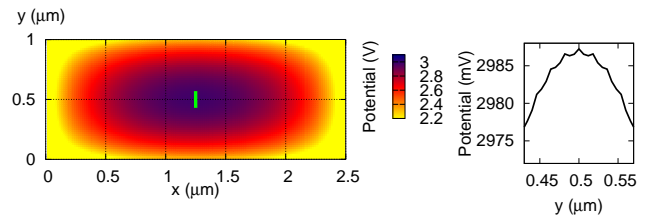}
\end{center}
\caption{Left panel (color online): potential $V(x,y,0)$ in the electronic subsystem S at $z=0$ due to the donor layer located at $z=-10$~nm. Right panel: cut through the donor potential in the middle of the bar (see bright line in the left panel at $x=1.25\;\mu$m) showing oscillations with amplitudes of 2~meV due to the regular donor lattice.
}\label{fig:DonorPotential}
\end{figure}

\section{Interacting particles simulation of semiconductor devices}
\label{sec:model}

Our first-principles model works in the spirit of molecular dynamics \cite{Aichinger2010a} by using microscopic equations for the motion of electrons in a charge-field and highlights the importance of electron-electron interactions together with the boundary conditions at particle reservoirs.
In the classical microscopic model, the force $F_k$ on the $k$th electron consists of the forces $F^{\rm C}_k$ due to all other $N=N_e+N_d$ electrons and donors, the confining potential $V_w$, and the velocity dependent Lorentz force $F^{\rm L}_k$,
\begin{eqnarray}\label{eq:forces}
\mathbf{F}_{k}&=&\mathbf{F}_{k}^{\rm C}+\mathbf{F}_{k}^{\rm L}-\nabla V_w(\mathbf{r})\big|_{\mathbf{r}=\mathbf{r}_k},\\\nonumber
\mathbf{F}_{k}^{\rm C}&=&\frac{q}{4\pi\epsilon_0 \epsilon} \sum_{\substack{l=1\\l\ne k}}^N
 \frac{q_l(\mathbf{r}_l-\mathbf{r}_k)}{|\mathbf{r}_l-\mathbf{r}_k|^3},\\\nonumber
\mathbf{F}_{k}^{\rm L}&=&q_k \dot{\mathbf{r}}_k\times\mathbf{B},
\end{eqnarray}
where $B$ denotes the magnetic field. Each particle state-vector contains the position $\mathbf{r}_k$, velocity $\dot{\mathbf{r}}_k$, and charge $q_k$ (negative for electrons, positive for the stationary donors).
In the following simulations we set the dielectric constant $\epsilon=8$ and use an effective mass $m^*=0.067\;m_e$ approximately matching electrons in a GaAs heterostructure.

Microscopic models of open systems present a recent development in theory \cite{Albareda2009a}. Only the computationally very demanding particle-based simulations are able to accurately treat electronic transport on the classical level and to go beyond simplified drift-diffusion or hydrodynamic approaches \cite{Vasileska2008a}. Our simulations of large systems ($10^3-10^4$ interacting classical particles) rely on a 200-1000 fold improvement of computational speed due to the use of high-performance graphics processing units, which  allow us to simulate micrometer-size devices at realistic electron densities. A quantum-mechanical calculation for such big open systems is still an impossible task. However, classical simulations have always provided guidance for future quantum-mechanical simulations and are needed for understanding the quantum-to-classical transition happening at higher temperatures. Especially for the understanding of the relation between classical and integer quantum Hall effect (IQHE) such simulations are required, since often for simplified quantum-mechanical calculations the Fermi-liquid approximation is used, where interactions are effectively not present between the quasi-particles. Notable exceptions are theories of the fractional quantum Hall effect (FQHE) and the analysis of small systems with Coulomb-blockade physics \cite{Albareda2009a}. Interestingly, electron-electron interactions are the corner-stone of the theory of the classical Hall effect and have to be considered in realistic semiconductor simulations at higher temperatures \cite{Vasileska2008a}. The particular shape of the classical Hall potential (Fig.~1) has also been observed in the IQHE (see Fig.~2 in Ref.~\onlinecite{Knott1995a}).

Also in graphene devices the absence of the integer and fractional QHE in a four-terminal measurement in small samples is attributed to the influence of the Ohmic contacts, demonstrating the need to include the finite geometry, the contacts, and the interactions in the device model \cite{Du2009a}. The theory of the IQHE in a finite graphene Hall device is discussed in Ref.~\onlinecite{Kramer2009b}.
Additionally, the IQHE displays a strong influence of the filling factor on the Hall potential, which is absent in classical models.

\section{Boundary conditions, contacts, and reservoirs in open systems}\label{sec:detail}

Our system Fig.~\ref{fig:System} consists of several connected parts: two metallic three-dimensional electron reservoirs R(source) and R(drain), two two-dimensional contact stripes C(source) over the complete device width between  $x=[0,0.1]$~$\mu$m, and C(drain) between $x=[2.4,2.5]$~$\mu$m where electrons from the reservoir are injected into a two-dimensional electron subsystem S. Uniformly distributed $N_d=8094$ positive background charges are located in the planar donor layer D $10$~nm beneath the electronic layer, whose extension exactly matches the extension of the electronic subsystem S. The resulting positive potential in the unfilled electronic layer (Fig.~\ref{fig:DonorPotential}) is bulged outward due to the finite extension of the system and has a range of $2.2$-$3.0$~V.
The motion of the electrons in the 2D subsystem is confined by a rectangular box-potential at $y=0$, $y=1$~$\mu$m, $x=0$, and $x=2.5$~$\mu$m, which represents the etched borders of the nanodevice. Perpendicular to the plane, a homogeneous magnetic field $B$ is present. 
The electrons sitting in the three-dimensional reservoirs are not part of our simulation and do not contribute to the potential calculation in the two-dimensional contact region, see Ref.~\onlinecite{Gonzalez1996a}, model M4. The contact points C$_i$ shown in Fig.~\ref{fig:LaplaceHallPotential} mark the positions of metallic spikes where electrons get transferred from the reservoirs to the subsystem S and vice versa.

To find the self-consistent stationary electron flow through the system and the corresponding potential, we put initially 7800 electrons at random positions in the electronic system S. The initial distribution of the electrons was chosen to counterbalance the positive donor charges, but we do not enforce equal numbers of electrons and donors in the system, since we want to obtain the self-consistent solution. 
The reservoirs are connected to the contact points and enforce an equipotential surface within the contact stripes C. In a typical Hall measurement the source and drain reservoirs and contacts are kept on fixed potentials and one probes the Hall-voltage in the transverse direction along the $y$-axis. In mesoscopic physics, interactions are often considered to be absent in the semi-infinite leads connecting the
device region with the reservoirs. In our setup, we find that the electron-electron interactions \textit{within the contact stripes}, from the moment of injection on, are crucial for the build-up of the self-consistent Hall-potential.
\begin{figure}[t]
\includegraphics[width=0.47\textwidth]{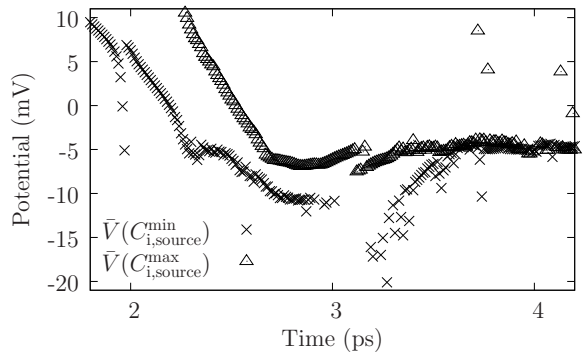}
\caption{Time-evolution of the extremal value of the potential of all contact points in the source contact towards the target value $V_{\rm source}=-5$~mV. The index $i$ in this graph does change with time and refers always to the location of the contact point $\mathbf{r}(C_{i,{\rm source}})$ with the greatest deviation from the target value.
\label{fig:Targets}}
\end{figure}
We allow the number of electrons within the subsystem to change due to injection and removal events in the contact stripes C. Several techniques have been proposed to establish a microscopic model of carrier injection from an Ohmic contact \cite{Gonzalez1996a,Vasileska2008a}. Our introduction of contact points is motivated by studies of the microscopic nature of Ohmic contacts \cite{Blank2007a}. Ohmic contacts to quasi two-dimensional electron gases (2DEG) in AlGaAs/GaAs heterostructures consist of metallic spikes which penetrate the cap-layer and reach down to the layer of the 2DEG \cite{Taylor1998a}. In one dimension, the most realistic results have been obtained by using the reservoir method, where electrons can enter the contact stripe from a reservoir in order to maintain the applied bias-voltage \textit{locally} in the contact stripe \cite{Gonzalez1996a}. We have generalized the 1D model to two dimensions and sample the potential in both contact stripes C at 500 contact-points C$_i$.
We specify the target value for the respective contact potentials (here $V_{\rm source}=-5$~mV, $V_{\rm drain}=+5$~mV). Initially, the electron and donor charges produce an on (spatial) average flat potential over the device region, which does in general not match the target potentials at the contacts. 

Next we integrate the equations of motion of the interacting $N$-body system in a magnetic field of $B=4$~T with a time-step of $\Delta t=5\times 10^{-17}$~s using a modified Euler algorithm
\begin{subequations}
\begin{align}
\dot{\mathbf{r}}_k(t + \Delta t) =& 
\frac{1-\omega_l^2 \Delta t^2}{1+\omega_l^2 \Delta t^2} \dot{\mathbf{r}}(t) +
\frac{2 \omega_l \Delta t}{1+\omega_l^2 \Delta t^2} 
\begin{pmatrix}
0 & -1 \\
1 & 0
\end{pmatrix}
\dot{\mathbf{r}}(t)\nonumber\\
&+\big[ \mathbf{F}_k^{\rm C}(\mathbf{r}_k(t))
-\nabla V_w(\mathbf{r})\big|_{\mathbf{r}=\mathbf{r}_k(t)}\big]/m_e
\\
\mathbf{r}_k(t+\Delta t) =& 
\mathbf{r}_k(t) + \Delta t~\dot{\mathbf{r}}_k(t + \Delta t),
\end{align}
\end{subequations}
where $\omega_l = \frac{e B}{2 m_e}$ denotes the Larmor frequency. During the propagation we keep track of the time-averaged potential at all contact points C$_i$
\begin{align}\label{eq:ContactPotential}
\bar{V}(C_i)=&\frac{1}{4\pi\epsilon_0 \epsilon \Delta T}\int_{t_0}^{t_0+\Delta T}  \rmd t
 \sum_{\substack{l=1\\l\ne k}}^N
 \frac{q_l}{|\mathbf{r}_l(t)-\mathbf{r}(C_i)|}\\\nonumber
&+V_w(\mathbf{r}(C_i)),
\end{align}
where $\mathbf{r}(C_i)$ denotes the position of the contact point $C_i$. The starting point of the integration is initially set to $t_0=0$.
The integration loop gets interrupted each $200 \Delta t$ and we determine for source and drain separately the contact point with the maximum value of the time-averaged potential $\bar{V}(C_{i,{\rm source}}^{\rm max})$, $\bar{V}(C_{i,{\rm drain}}^{\rm max})$ and the point with the minimum value $\bar{V}(C_{i,{\rm source}}^{\rm min})$, $\bar{V}(C_{i,{\rm drain}}^{\rm min})$. However, we discard the selected points in case the integration time is smaller than a cyclotron period $\Delta T<T_c$ in order to average out momentary potential fluctuations.
Next we inject an electron at location $\mathbf{r}(C_{i,{\rm source}}^{\rm max})$ in case $\bar{V}(C_{i,{\rm source}}^{\rm max})>V_{\rm source}$ and we mark the electron closest to $\mathbf{r}(C_{i,{\rm source}}^{\rm min})$ for removal if $\bar{V}(C_{i,{\rm source}}^{\rm min})<V_{\rm source}$. The same process takes place at the drain contact.
The rate of 4 possible removals and additions every $200 \Delta t$ corresponds to a maximum possible current of $16\;\mu$A. After an injection or removal event, the lower limit $t_0$ of the time-integration in Eq.~\ref{eq:ContactPotential} is set to the time of event.
In our simulations we obtain currents of the order of $1\;\mu$A, demonstrating that convergence was achieved and that there is no need to inject or remove electrons every $200 \Delta t$ in order to maintain equipotential surfaces in the contact regions. Between injection and removal, the electrons move under the forces given by Eq.~(\ref{eq:forces}).

The specific advantage of the used NVIDIA Tesla GPU board lies in the simultaneous calculation of 240 two-body interaction forces, which yields performances in the teraFLOPS ($10^{12}$ floating point operations per second) range \cite{Nyland2007a}. The limitation to single-precision arithmetic is of no major concern for the highly-chaotic classical simulation, since we are interested in statistical averages.
We perform 200 iterations steps on the GPU, before we transfer the instantaneous positions and velocities of all electrons to the CPU host. On the CPU, we calculate the potentials in the contact regions and inject and remove particles according to the rules given above. Subsequently the updated set of particles is sent back to the GPU for propagating another 200 steps. The technical details of our hybrid GPU-CPU scheme will be described elsewhere.

Finally, we calculate the resulting (time-averaged) potential at all points of the electron subsystem S. Our injection and removal procedure ensures that the potentials at the source and drain contacts converge towards their respective target values (see Fig.~\ref{fig:Targets}), which are physically enforced by the presence of the metallic particle reservoirs. If instead electrons are injected with equal probability across the contact stripe, no Hall voltage across the device emerges.
\begin{figure}[t]
\includegraphics[width=0.47\textwidth]{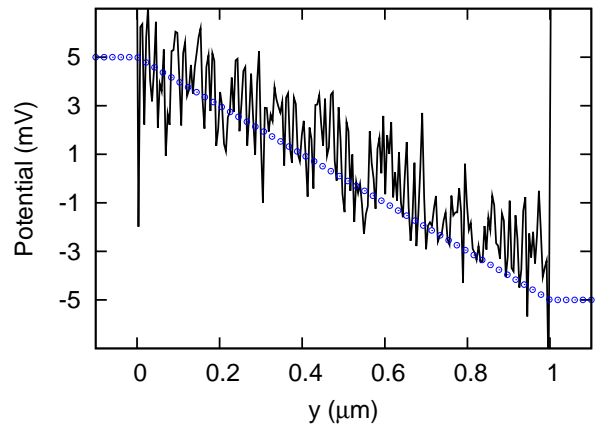}
\caption{(color online) Cut through the time-averaged potential across the center of the Hall bar at $x=1.25\;\mu$m. The target potentials are $+5$~mV (left source contact) and $-5$~mV (right drain contact). The oscillations in the potential profile are caused by the oscillations of the donor potential, shown in the right panel of Fig.~\ref{fig:DonorPotential}, the blue circles denote the conformal map solution. The potential map of the complete Hall bar is given in Fig.~\ref{fig:AveragePotential}. 
\label{fig:AveragePotentialSlice}}
\end{figure}

\section{Time-averaged potentials and role of interactions}\label{sec:result}

\begin{figure}[b]
\begin{center}
\includegraphics[width=0.47\textwidth]{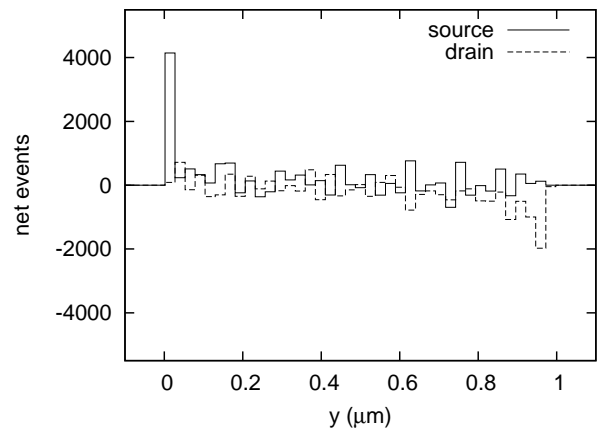}
\end{center}
\caption{Spatial distribution of the net injection (positive axis) and removal (negative axis) events after a total of $10^5$ events across the source and drain contacts. Electrons are predominantly injected at the lower left corner of the source contact, and are most often removed at the upper right corner of the drain contact.
}\label{fig:Histogram}
\end{figure}

\begin{figure}[t]
\subfigure[]{
\includegraphics[width=0.47\textwidth]{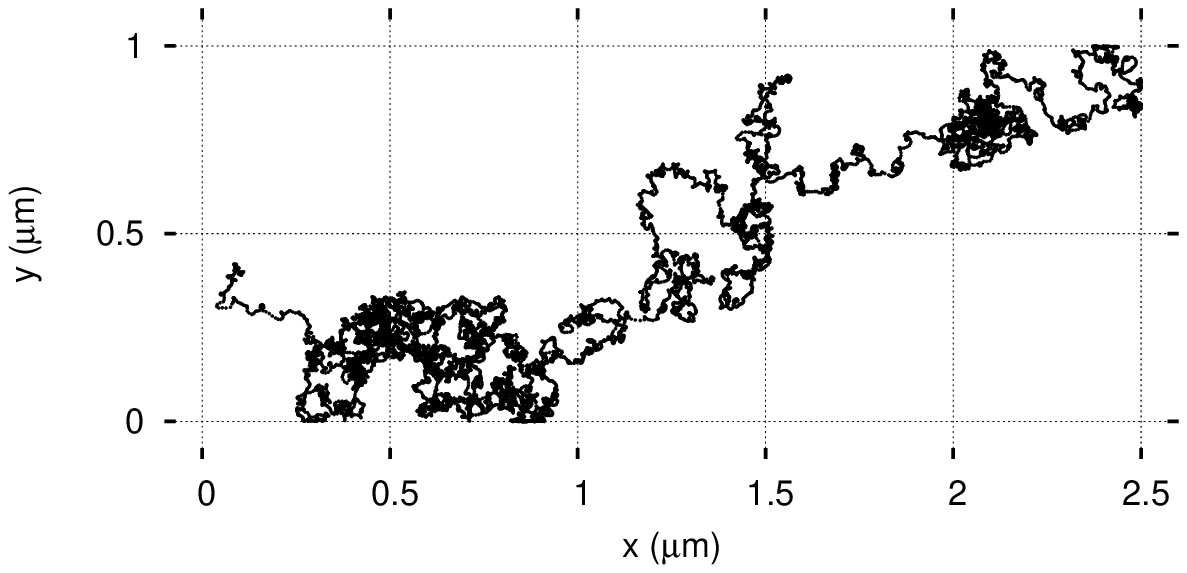}\label{fig:TypicalPath}
}
\subfigure[]{
\includegraphics[width=0.47\textwidth]{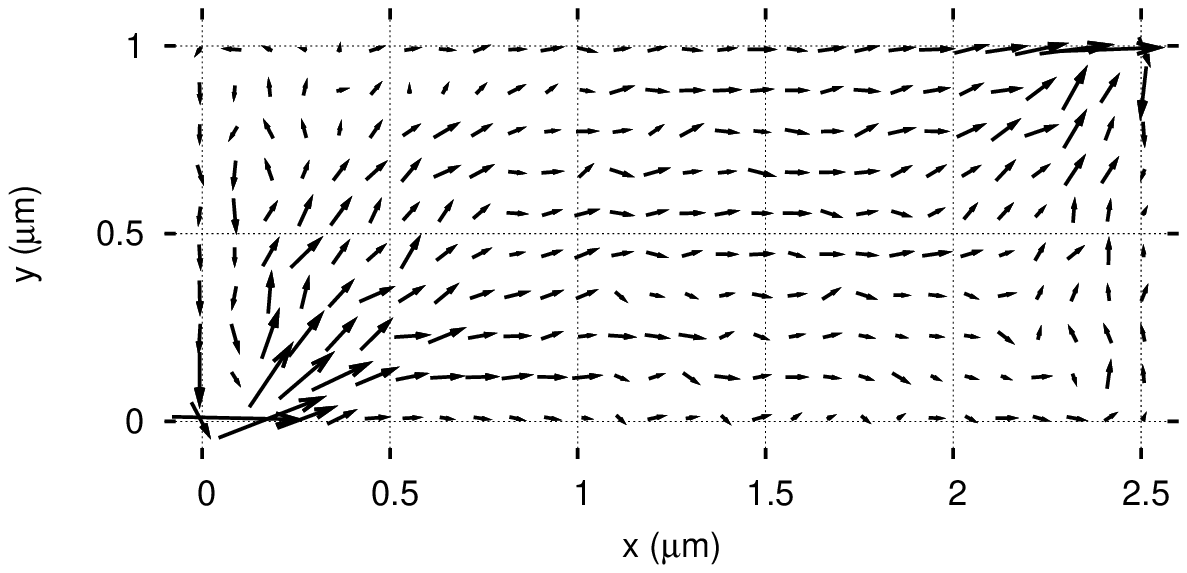}\label{fig:CurrentDensity}
}
\caption{(a) Typical source-drain trajectory starting at the left source contact and reaching the right drain contact. 
(b) Time-averaged current density distribution. The arrow length is proportional to the local magnitude of the current density.}
\end{figure}

\begin{figure}[b]
\begin{center}
\includegraphics[width=0.47\textwidth]{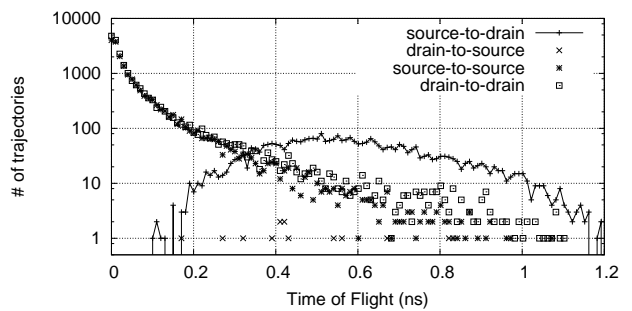}
\end{center}
\caption{Histogram of the time-of-flights of the injected electrons: number of source-to-drain paths (solid line), drain-to-source paths (dashed line), source-to-source paths (dotted line), and drain-to-drain paths (dash-dotted line). For comparison, the fastest mean-field drift path yields a time-of-flight of $0.7$~ns.}\label{fig:tof}
\end{figure}

We track the total number and the position of the electrons in the Hall bar. After a transient behavior (lasting ca.\ 0.1~ns) at the beginning of the simulation, a steady-state situation is reached where the electron number fluctuates around the value $N_e=7975\pm5$, giving rise to an average electron density of $n_e=3.19\times 10^{15}$~m$^{-2}$.
The current is obtained by introducing a flux-surface across the subsystem S or alternatively by counting the injection and removal events at each contact \cite{Vasileska2008a}. The net-current inflow from the source contact is $+1.07\;\mu$A, while the drain contact supports a net outflow of $-1.07\;\mu$A. The expected Hall voltage $V_{xy}=R_{xy}\;I$ across the device is related to $n_e$ via
\begin{equation}
R_{xy}=\frac{B}{e n_e}=7826\;\Omega.
\end{equation}
The calculated Hall voltage $V_{xy}=8.4$~mV is in good agreement with the target potential difference of $10$~mV and the time-averaged value (see Fig.~\ref{fig:AveragePotentialSlice}).
Figure~\ref{fig:Histogram} shows the histogram of the injection and removal events in the contact stripes, where for each bin we display only the net-result (number of injection events minus removal events) out of $10^5$ events. The source contact predominantly injects electrons in the lower left corner, while the drain contact removes most electrons in the upper right corner. Thus the mean electron flow  follows a diagonal path across the device as displayed in Fig.~\ref{fig:CurrentDensity}. The trajectory of an individual electron (Fig.~\ref{fig:TypicalPath}) can deviate considerably from the average behavior. Figure~\ref{fig:tof} shows the distribution of the flight times of the electrons between the contacts. 
From the time-of-flight distribution we infer that most electrons are short lived and get absorbed within the same contact in which they got injected. These electrons do not contribute to the total current, but they are important to establish the mean Hall potential and the average current-density field. The average time-of-flight from source to drain contact is ${\bar{t}_{\rm tof}}=0.6$~ns, and sets the time-scale required in order to obtain the converged result shown in Fig.~\ref{fig:AveragePotential}. Starting the time-average at a later point $t>{\bar{t}_{\rm tof}}$ does not alter the picture.

We also studied the importance of the electron-electron interactions by changing the dielectric constant $\epsilon$ in a range from $1-8$ and by additionally introducing a short-range cutoff of the Coulomb interactions. We find that for decreased Coulomb interactions ($\epsilon>10$) the resulting potential does not resemble the conformal map result, but instead electrons start to pile up next to the converged contact region, while at the other side the electron density is reduced and results in a too positive value of the potential. Our finding shows that the electron must have a minimum degree of incompressibility in order to yield the classical Hall effect. Antisymmetrization and the Pauli principle (which are not part of the classical simulation) can provide other mechanisms to keep electrons apart and give an effective ``Pauli incompressibility''.

\section{Connection between classical and quantum Hall effect}\label{sec:boundary}

As exemplified by our particle-simulation of a Hall-bar, the incorporation of the metallic boundary conditions together with the electron-electron interactions in the whole device (including the contact stripes) and the finite size of the system are all mandatory ingredients for the formation of a classical Hall potential. These facts were already noted by Hall in 1879 \cite{Hall1879a}, p.~287, but the precise way how these ingredients lead to the Hall effect could not been elucidated before due to the lack of computational methods.

An intriguing problem is the passage from the classical to the IQHE and the role of interactions in the IQHE. Recently it has been suggested that interactions are actually an important ingredient for the IQHE in order to calculate the critical exponents of percolation theories of the IQHE \cite{Slevin2009a}. The experimental observation of the ``classical'' Hall potential \cite{Knott1995a} under conditions where the resistivity is quantized provides another hint that electron interactions and the boundary conditions at the reservoir/subsystem interface plays an important role in the IQHE.
Interactions enter the picture on two different length scales: while the Coulomb repulsion, the specific boundary conditions set by the device geometry, and the metallic contacts are seen to enforce the global shape of the Hall potential, small clusters of neighboring electrons perform a highly correlated dancing pattern, which could be seen as the classical analogue to the FQHE \cite{Wen1995a}.
The interactions between the donors and the electrons result in the screening of the positive donor charges, which become visible after performing the time-average of the instantaneous potential snapshots.

A direct translation of the classical calculations to the quantum regime is not easily possible, since in the quantum-case at low temperatures scattering events strongly depend on the occupation of initial states and the availability of final ones. Furthermore exchange and correlation effects have to be taken into account \cite{Filinov2009a} and the Coulomb interaction is modified due to the orbital extension of the electronic Landau levels. As a first step towards the incorporation of quantum-effects, we have repeated the simulations using a classical Hartree-approach, where Gaussian-shaped charge-clouds represent the density of an electron within the first Landau level. The Gaussian distribution cuts off the Coulomb-interactions at short distances, but does still lead to the buildup of the Hall potential shown in Fig.~\ref{fig:AveragePotential}. In all our classical calculation, skipping orbits play only a minor role for the transport and the frequent collisions with other electrons lead to a detachment of the trajectories from the edges. 

This observation poses the question whether in an interacting quantum-mechanical many-body calculation edge-states prevail. The electric Hall field and dissipation at the contacts increase with increasing current and inelastic scattering events may provide another mechanism to change from an edge-transport picture to bulk transport diagonally across the device. The experimental observations of the Hall potential similar to the one shown in Fig.~\ref{fig:LaplaceHallPotential} on a QHE plateau demonstrates that a quantized resistivity can occur hand-in-hand with the diagonal transport picture. Transport along the edges is in our classical model not compatible with the equipotential boundary conditions at the contacts, since the arrival of edge current in the contact would increase the electron density locally and lead to deviations of the potential from the prescribed value. If in the quantum-mechanical case a self-consistent calculation \cite{Poetz1989a} can remedy this situation requires further investigations.

\section{Conclusions and outlook}\label{sec:conclusion}

In conclusion, we have developed a highly efficient classical particle-based simulation scheme for semiconductor devices, which incorporates the spatial and temporal sequence of injection and removal events happening at Ohmic contacts. The classical Hall potential did emerge from our calculation as the time-averaged self-consistent potential. The choice of local boundary conditions within the metallic contacts has a crucial influence on the resulting global particle distribution inside a Hall bar.
Our method extends previous models\cite{Vasileska2008a,Albareda2009a} and can be readily adapted to  include charged gates and different device geometries. We expect that our new computational method opens a window towards a first-principles treatment of interaction effects in semiconductor devices and accelerates the development of the next generation of transport codes for realistic device settings.

\subsection*{Acknowledgements}

We appreciate stimulating discussions with P.~Kramer, C.~Kreisbeck, E.~R\"as\"anen, J.~Fabian, K.~Richter, and D.~Weiss. We would like to thank the Harvard SEAS IT department for GPU computing time. TK and VK are funded by the Emmy-Noether program of the DFG, grant KR 2889/2-1.

\providecommand{\url}[1]{#1}

\end{document}